\documentstyle[12pt,aasms4]{article}

\slugcomment{Submitted to the Astronomical Journal}

\lefthead{Keel \& White}
\righthead{HST Imaging of Overlapping Galaxies}

\begin{document}

\title{Seeing Galaxies Through Thick \& Thin. III.\\
HST Imaging of the Dust in Backlit Spiral Galaxies\altaffilmark{1}}

\author{William C. Keel\altaffilmark{2} and Raymond E. White III\altaffilmark{2,3}}


\altaffiltext{1}{Based on observations with the NASA/ESA {\it Hubble Space Telescope}
obtained at the Space Telescope Science Institute, which is operated
by the Association of Universities for Research in Astronomy, Inc.,
under NASA contract No. NAS5-26555.} 
\altaffiltext{2}{Department of Physics and Astronomy, Box 870324, University of Alabama,
    Tuscaloosa, AL 35487}
\altaffiltext{3}{Code 662, NASA Goddard Space Flight Center, Greenbelt, MD 20771}


\begin{abstract}
We present analysis of WFPC2 imaging of two spiral galaxies partially
backlit by E/S0 systems in the pairs AM1316-241 and AM0500-620,
as well as the (probably spiral) foreground system in NGC 1275.
Images in $B$ and $I$ are used to determine the reddening curve of 
dust in these systems. The foreground spiral component of AM1316-241 shows dust
strongly concentrated in discrete arms, with a reddening law
very close to the Milky Way mean ($R=E_{B-V}/A_V= 3.4 \pm 0.2$). The dust
distribution is scale-free between about 100 pc and the
arm dimension, about 8 kpc. The foreground spiral in AM0500-620
shows dust concentrated in arms and interarm spurs, with measurable
interarm extinction as well. In this case, although the dust
properties are less well-determined than in AM1316-241, we find
evidence for a steeper extinction law than the Milky Way mean
(formally, $R \approx 2.5 \pm 0.4$, with substantial variation depending on
data quality in each region).
The shape of the reddening law suggests that, at least in AM1316-241, we have 
resolved most of the dust structure. In AM0500-620 it is less
clear that we have resolved most of the dust structure, 
since the errors are larger. In AM0500-620, the slope of the
perimeter-scale relation (associated with fractal analysis)
steepens systematically on going from regions of low to high extinction. 
A perimeter-smoothing length test for scale-free (fractal) behavior in 
AM1316-241 shows a logarithmic slope typically 
$-0.4$ on 100-1000 pc scales. However, we cannot determine a unique fractal
dimension from the defining area-perimeter relation, so the projected dust 
distribution is best defined as fractal-like. For scales above 2-4 pixels 
(120-250 pc), the box-counting estimate gives a fractal
dimension close to 1.4, but the perimeter-area relation gives a dimension
of 0.7 on large scales and inconsistent results for small scales, so that
the distribution shows only some aspects of a fractal nature.
In neither galaxy do we see significant regions even on single-pixel scales 
in spiral arms with $A_B > 2.5$. The measurements in NGC 1275 are compromised 
by our lack of independent knowledge of the foreground system's light
distribution, but masked sampling of the absorption suggests an effective
reddening curve much flatter than the Milky Way mean (but this may 
indicate that the foreground system has been affected by immersion
in the hot intracluster gas, or is inside the stellar
distribution of NGC 1275). The bright blue star clusters trace the absorption
in this system quite closely, indicating that these clusters belong to
the foreground system and not to NGC 1275 itself.

\end{abstract}




%

\section{Introduction}

The importance of dust extinction within galaxies is a fundamental
problem for a wide range of issues. It affects the extragalactic
distance scale through inclination corrections to the Tully-Fisher relation. 
Dust extinction affects the inferred evolution of quasars (the high-redshift 
cutoff, in particular) via cumulative extinction by foreground
galaxies along typical lines of sight. 
Estimates of star-formation rates and mass-to-light ratios in 
individual galaxies are strongly dependent on estimates of internal
absorption in galaxies. 
Assessing the ``typical" extinction within spiral galaxies
has proven controversial, and various approaches give substantially 
different results (and the various interpretations of these results diverge 
even more widely). Although the presence of obscuring matter was
clear from the earliest systematic galaxy photography (\cite{curtis18}),
only much later was the connection between inclination and
surface brightness used to estimate how much internal extinction
there might be (\cite{holmberg58}). Holmberg's reassuring conclusion, that 
internal extinction is not a major factor in the global emerging 
light from spiral galaxies, eventually came to be challenged on two grounds.
Disney et al. (1989) showed that the blue colors and inclination behavior
of spiral disks could be consistent with models of high optical depth, so that 
what we see is dominated by a small proportion of unreddened starlight. 
Valentijn (1990) drew attention to this issue by analyzing the surface brightness
and inclination of spirals in the ESO-LV catalog (\cite{esolv}), and claiming
that these quantities are empirically so independent that typical spirals
must be optically thick to blue light across much of their visible disks.
This analysis (though not necessarily its conclusion) was challenged by
Burstein et al. (1991), who attributed Valentijn's main result to possible
selection effects in the galaxies included in the ESO-LV sample, and asserted
that the problem could only be solved when truly volume-limited galaxy
samples could be assembled.

Because of the assumptions required in using these model-based techniques
to estimate extinctions, we have pursued a more direct approach:
observing overlapping galaxy pairs to measure dust extinction in foreground
spirals via differential surface photometry. 
As set out in White \& Keel (1992), the ideal case for
such an analysis has a foreground spiral
half projected against a similarly sized background elliptical. This geometry allows
us to recover estimates of the light from each galaxy by itself,
using symmetry from the non-overlapping portions of the images, and
thence to reconstruct the absorption in the overlapping part of the
foreground spiral, limited only by the level of asymmetry in the spiral
(since elliptical and S0 systems are so precisely symmetric). An extensive
ground-based program (\cite{wkc}, hereafter WKC; \cite{wkc96}; \cite{kw}) 
identified the best
candidates for such analysis, and measured disk opacities for 10 spiral
galaxies. The general results were:

-- spiral arms and resonance rings can have large optical depths 
($\tau_B > 2$) over a wide range of galactocentric distance.

-- interarm dust has a roughly exponential distribution with radius,
and causes only mild extinction in the outer disk (typically $A_B \geq 0.5$ 
only within 0.4$R_{25}$).

-- the extinction ratios from $B$ to $I$ passbands are greyer than the
mean Milky Way curve, which we interpreted as evidence of patchy
or clumpy extinction on unresolved size scales.

This last point made it especially desirable to measure dust properties on
finer spatial scales, perhaps recovering the intrinsic reddening curve of
the material and facilitating a comparison between grain populations in
various spiral galaxies. This was pursued using two approaches. We describe
here high spatial resolution observations achieved by HST imaging of two 
especially amenable galaxy pairs, AM1316-241 and AM0500-620 (plus archival data
on the backlit high-velocity system in NGC 1275). 
A complementary analysis, comparing far-IR and submillimeter emission
with optical absorption measures for three overlapping galaxy pairs
(including AM1316-241) was presented by Domingue et al. (1999).

These galaxy pairs were selected as the most promising cases, for symmetry 
and angular size, from the initial survey reported by WKC. The two newly-observed 
pairs were both initially catalogued
by Arp \& Madore (1987). AM1316-241 (ESO 508-IG45) consists of a highly
inclined spiral in front of an elliptical galaxy with very nearly circular 
isophotes, as measured from the nonoverlapping regions.
The spiral has measured redshift $cz=10365$ km s$^{-1}$, notably
larger than that of the elliptical (9700 km s$^{-1}$, White \& Keel
1992; Donzelli \& Pastoriza 2000). AM0500-620 (ESO 119-IG27) consists of a 
nearly face-on spiral
of type Sb partially projected in front of a smooth system with a light
profile suggesting an S0 classification. 
Donzelli \& Pastoriza list heliocentric velocities of 9005 and 8811 km s$^{-1}$
for the spiral and elliptical components, respectively. For linear scales, we 
adopt a Hubble constant of 75 km s$^{-1}$ Mpc$^{-1}$, for which a single WFPC2
$0.1\arcsec$ pixel subtends 65 and 57 pc, respectively, in AM1316-241 and AM0500-620. 
We also consider archival WFPC2 data on the object in the foreground of 
NGC 1275; in this case the foreground system has a redshift 
$cz=8200$ km s$^{-1}$, while NGC 1275 
itself has $cz=5264$ km s$^{-1}$. Given their context in
the Perseus/Abell 426 cluster, 
the line-of-sight separation of these two galaxies is almost indeterminate.
Some suggest that the two galaxies are strongly interacting,
perhaps enhancing the luminosity of the emission-line filaments around
NGC 1275 proper (\cite{hu83}), and absorption geometry indicates that the 
foreground system  must lie at least outside these filaments 
(\cite{keel83}).
For the PC CCD, which encompasses most of the foreground absorption, 
the image scale in this case is 15 pc/pixel, although we worked
with mosaiced data at 30 pc/pixel.

\section{Data and image analysis}

The two overlapping elliptical/spiral pairs AM0500-620 and AM1316-241 were
observed with WFPC2 in the $B$ (F450W) and $I$ (F814W) passbands,
with rootnames for the image sequences of {\tt u3lw04} 
and {\tt u3lw03} respectively.
In each case, the total exposures were 2000 seconds ($I$) and 2600 seconds ($B$),
split into halves for cosmic-ray recognition and rejection. 
The individual exposures in each filter were combined using the {\tt crrej} 
task in STSDAS for
cosmic-ray removal, plus cleaning ``by hand" of a few particle
events missed by the automated procedure.
Additional
120-second $B$ and 60-second $I$ exposures were taken in case the galaxy nuclei 
saturated on the longer exposures, as indeed happened in $I$ for a 
$1.0 \times 0.5$-arcsecond region at the center of the elliptical in AM1316-241.
Both members of AM0500-620 fit comfortably in the WF3 CCD field, while
AM1316-241 stretched across both WF2 and WF3, with WF2 containing the
outer end of the foreground disk. Images in both filters were registered
to within 0.2 pixels, so we compare the two passbands without
any pixel interpolation.

As often happens, the WFPC2 images reveal significant structures in the 
background galaxies which were not at all
apparent from our ground-based imagery. Fortunately, these have at most
a slight impact on our ability to do the image modeling required
for an opacity measurement. The background galaxy in AM0500-620 
(Fig.~1) now
shows a central spiral pattern spanning the innermost $2.6\arcsec=1.4$ kpc,
including dust lanes and bright knots (of the appropriate luminosity to be 
star clusters, at $B=20.3$ or $M_B=-15$ and fainter). We therefore cannot 
measure foreground absorption this close to the nucleus, since
it is unclear which galaxy is associated with a particular small dust feature. 
The elliptical in AM1316-241 (Fig.~2) has a small linear bright 
feature, perhaps an edge-on stellar disk, spanning a diameter of $1.3\arcsec=0.8$ kpc
and aligned in roughly the same direction as the foreground spiral disk.
This structure has no effect on our measurements of dust in the foreground 
galaxy.

As described by WKC, our opacity measurements 
require modeling of the intrinsic light distributions of both foreground
and background galaxies in a region where they overlap. 
In such a region of overlap, if the background galaxy has an unabsorbed intensity 
$I_B$, the foreground galaxy contributes an intensity $I_F$,
and the total observed intensity is $I$, then the effective
optical depth $\tau$ through the foreground galaxy is
$$ e^{- \tau} = {{ I - I_F} \over {I_B}} .$$
In general, we of course cannot directly separate foreground from background
emission in regions where the two galaxies are superposed.
We instead estimate $I_F$ and $I_B$ from
symmetric, non-overlapping regions of the galaxies.
The two pairs we consider here were selected
to be the best pairs known for such an analysis, based on angular
size and symmetry of both pair members and the improved modeling that
is possible if each galaxy is about half overlapping, leaving the
other half as a symmetry template. For background E/S0 galaxies,
estimates of $I_B$ are especially accurate (to better than 1\% ) since
we can make a two-dimensional symmetric model based on all parts of the galaxy
that are unlikely to be overlapped by any material associated with the
foreground galaxies. In practice, stars, background galaxies, and any other
fine structures are masked during model fitting. Residuals from the fit 
can be evaluated in the non-overlapping areas, giving us a point-by-point
assessment of how accurate the modeling has been. We perform these 
operations in the observed pixel frame, without any rebinning which
would degrade resolution in some parts of the field of view. 

In the case of NGC 1275, the foreground system and intrinsic structures
cover enough area that
several rounds of masking were needed to get a smooth symmetric background
model. We used symmetry to provide an initial guess at intensities in
the area not imaged around the PC chip, as needed by the STSDAS {\it ellipse}
routine to provide a large enough fraction of the elliptical isophotes
in some ranges of semimajor axis.

Larger errors will be introduced through uncertainties in the foreground
galaxy intensity $I_F$, since spirals
are rich in structure. AM0500-620 is attractive for this analysis since
its grand-design pattern is highly symmetric; the ridgelines of its
spiral pattern lie within $1\arcsec$ of their mirror reflections. Our point of
departure for modeling it was rotation through $180^\circ$, median
filtering with an 11-pixel ($1.1\arcsec$) square window to remove bright clusters
and associations so we need to deal only with positive structural
residuals, and subtraction.  
In AM1316-241, the
opposing outer arm at the relevant radius is both smooth and of rather
low surface brightness, so that it does not offer a large uncertainty
in accounting for foreground light if the spiral is even roughly symmetric
at these radii. It proved useful to model the galaxies 
in each pair iteratively,
using a first-guess model of the background galaxy to isolate
emission from the foreground galaxy, subtract this from the data and
derive an improved background model, and so on. Scatter in the residuals
indicates that the errors in $e^{-\tau}$ are 
0.11-0.15 per pixel in $I$ for AM0500-620
and somewhat better, 0.06-0.10, for AM1316-241; photon statistics
and systematic errors from structure in the foreground galaxies
contribute about equally to the error budget.
The foreground corrections range from 5--30\% in AM1316-241
and from 2--15\% in AM0500-620, so that departures from symmetry in the
foreground light distribution will be multiplied by these factors
in their impact on derived opacity values.

Scattering of background galaxy light by dust in the foreground galaxy 
could reduce the effective optical
depth we measure. Numerical integrations to estimate this effect,
starting from plausible line-of-sight separations between the galaxies,
were described by WKC, and these remain valid for the new data.
The relevant quantity is not the fractional amount of scattered
light itself, but the differential scattering between the point
under consideration for opacity measurement and its symmetric counterpart
in the foreground disk which was used
to estimate $I_F$. For these pairs, this effect
is small: less than 3\% for AM0500-620 and less than 1\% for AM1316-241.

Realization of Eq.~(1) for these images yields maps of residual intensity
$e^{- \tau}$ as shown in Figs.~3--5. We work primarily
with residual intensity
(transmission), denoted $T_B$ and $T_I$ 
rather than optical depth $\tau$ or magnitudes of extinction $A_B$ and $A_I$ for 
statistical measurements,
because the errors in this linear quantity are
much better behaved (being symmetrically
distributed) than in the logarithmic measures. In both systems, the
$I$ data are of higher quality than those in $B$ despite the smaller
extinction at $I$, resulting both from the better $S/N$ ratio of
the initial data, the better symmetry, and reduced influences of
young clusters and associations in the redder passband.

\section{Distribution of dust absorption}

The spiral components of these pairs offer interesting distinctions,
suggesting some of the variety that must be present among spiral galaxies.
The dust in AM1316-241 is very strongly concentrated to the prominent
arm, now seen as a rich filamentary pattern, with none seen outside this arm
(to quite low limits) and only a small amount inside the arm (until the
background intensity has dropped too low to allow measurement of dust
in the next spiral feature inward). In AM0500-620, there are several 
very narrow spiral features outlined by dust, including a spur between
two of the prominent arms, plus a significant component of interarm
obscuration. The arm/interarm distinction that proved useful
in ground-based work remains valid at this increased
resolution, and is in fact even clearer. At all
projected radii that we can sample, we see a strong distinction 
between dust in spiral arms
and spurs on the one hand, and more smoothly distributed interarm dust on 
the other. The regions we can usefully analyze span the range
$0.53-0.72 ~ R_{25}$ in AM0500-620, and $0.44-0.80 ~ R_{25}$ in AM1316-241.

One global description of absorption effects is the fraction of area covered
at various extinction values. Fig.~6 shows both differential and cumulative
distributions of area as functions of transmission $T_B$ for the regions in AM1316-241 
and AM0500-620 which are well measured.
The two are different in detail. In particular, 
AM0500-620 has a larger fraction of area with low opacity than AM1316-241;
furthermore, AM1316-241 has an obviously bimodal distribution of opacity, 
while AM0500-620 does not.
Half the area in AM1316-241
is covered by dust with a $B$ transmission of $T_B=0.63$ or greater,
while the corresponding halfway point in AM0500-620 is $T_B\ge0.82$.
Typical pixel-by-pixel errors are 0.1 in transmission, as shown
by the width of the spike near zero extinction ($T_B=1$) in AM1316-241, 
so that the number of points with $T_B<0.1$ is consistent 
with scatter from points higher in the distribution, so that
we see no evidence for pixel-scale areas passing less than 10\%
of the light ($A_B > 2.5$). These distributions are clearly area-
rather than intensity-weighted, so these are the kinds of measures
that are useful in coverage problems such as evaluating the effect
of galaxy disks in absorbing QSO light.

We use the compromise dust mixture from Domingue et al. (1999) to
estimate the dust mass seen in these regions, and compare arm and
interarm contributions. This mixture (about midway between 
pure graphite and pure silicate grains in their
optical properties and density) has
mass column density $\rho = 2.41 \times 10^{-5} \tau_B$ g cm$^{-2}$,
or $0.12~\tau_B$ $M_{\sun}$ pc$^{-2}$. The transmission maps can thus
be turned into column density maps for each overlapped area. As
a guide in interpreting the figures, $\tau=1$ in the $I$ images
corresponds to a dust mass per pixel of 900 $M_{\sun}$ in AM0500-620 and 
1166 $M_{\sun}$ in AM1316-241.

In AM0500-620, the arm/interarm contrast is much higher than we
saw from the ground in WKC, because much of what appeared to
be diffuse interarm extinction is now resolved into the prominent
spur cutting across the main spiral pattern. Defining arm features
at $\tau_B \ge 0.15$, the boundary of contiguous arm extinction, or
$57$ $M_{\sun}$ per pixel, we find that 97\% of the dust mass is in
arm regions comprising 54\% of the projected area. Higher column-density
cuts yield further relations: 55\% (36\%) of the dust mass is contained
within 20\% (10\%) of the projected area.
The most prominent cloud complex,
above center in the spur, has a total dust mass of about $1.4 \times 10^5$
$M_{\sun}$. For a typical Galactic gas-to-dust ratio of 300 by mass,
this implies a total mass in the cloud of order $4 \times 10^7$ $M_{\sun}$,
typical for a large molecular-cloud complex extending almost
2 kpc in an arm. 

Individual clouds are less distinct in AM1316-241.
Several individual clumps on $0.4-0.8\arcsec$ (250-600 pc) scales have dust masses
in the range $3-7 \times 10^4$ $M_{\sun}$. The concentration to the
arm pattern is pronounced here as well, with 89\% of the mass
in the overlap region contained in contiguous structures
occupying half the area. The 10\% of the area with the highest
column densities contains 31\% of the dust.

Our analysis for NGC 1275 is necessarily less complete than for the two AM
pairs, since photometry alone cannot be used
to independently assess the amount of
foreground light coming from a completely backlit galaxy.
We have tried to limit the impact of this uncertainty by masking all pixels
bluer than a limiting value in $B/I$ flux ratio, which includes
many pixels with obvious absorption. This leaves us unable to derive
reliable covering fractions, since the blue clusters are distributed in
almost the same way as the dust (frequently with a systematic offset
such as is often seen between dust and stars in typical spiral arms).
In fact, the distributions of blue clusters and absorption (Fig.~5)
suggest that most of the bright blue clusters studied by Carlson et al. (1998)
are in fact part of the foreground galaxy and not of NGC 1275 itself.

Of the regions we can measure in NGC 1275, $T_I=0.45$ is the lowest $I$-band
(F702W for this object) transmission found in areas large enough to
be real detections; the areal distribution of $T_I$ is
nearly flat from $T_I=0.60-0.90$. The dust lanes are about as optically
thick as the spiral arms in the other two systems.
However, as noted below, the extinction curve in this system is
drastically different from what we find for the two other spirals.
This may signify a real difference in the grain population,
or it may be due to the foreground galaxy being embedded in the ``background'' 
galaxy, so some of the near side light from  the ``background'' galaxy may not be 
absorbed by the foreground galaxy. 
The structures of both absorption and excess
emission from blue clusters and associations strongly resemble
a late-type spiral in at least the northwest half of the foreground
system, as seen in Fig.~5. This is in accord with several previous
studies identifying this object as a late-type spiral, seen almost
edge-on and perhaps being disrupted by a very rapid tidal encounter with the
background galaxy (\cite{keel83}; \cite{hu83}).

\subsection{Fractal structure in dust distributions?}

Indications that the H~I in several nearby galaxies can be modeled
as scale-free fractal distributions (Westpfahl et al.~1999) motivate a
similar analysis of the dust maps described above. 
One prominent symptom of fractal behavior is
a scale-free relation between the perimeter of a contour and the scale 
length over which it is integrated or smoothed for the measurement.
We employ a perimeter-scale test,
implementing the box-counting procedure of Westpfahl et al. 
Specifically, the length of various contours was evaluated in
transmission maps for various values of $T_I$ and
for smoothing lengths of 1, 2, 4, 8, and 16 pixels, so that no resampling
of pixel values affects the results. This evaluation was done by creating
masked images cut above and below desired thresholds, then counting the
number of pixels making the transition. We plot this relation using
perimeters as scaled to the original pixel size (rather than
block-averaged effective pixels at each scale), so the box-counting
dimensions will be the slope of the relation plus one. 

In AM1316-241, the perimeter-scale relation is nearly linear 
(in the log for both quantities) over scales from 2--16 pixels
(130--1000 pc), when contours are used for transmission values $T_I=0.5-0.9$
on the higher-quality $I$ map (Fig.~7). The 
measured perimeter for one-pixel smoothing
scales (i.e. the original data) is larger than the extrapolated
value from larger smoothing lengths, for all transmission
levels. Numerical experiments using block-averaged and PSF-convolved
versions of the original image confirm that this discrepancy cannot
be due to the numerical properties of the algorithm (which operates
stepwise in exactly the same way on going from 1--2 pixels as, for
example, from 2--4);
this result is in the opposite sense of any smoothing effects due to the finite 
PSF of WFPC2 images (which would
contribute about a 20\% perimeter reduction 
due to smoothing). However, the effects of finite signal-to-noise ratio
in the data can contribute significantly to apparent structure at
small opacities on scales 1--2 pixels. The effect is most apparent
for small opacity because the local slope in residual intensity is
smaller, so a given amount of noise will move the contour at a particular
level by a large amount. We estimate the magnitude of the effect by recomputing
the area-perimeter relation for smoothing lengths of 1,2, and 4 pixels for
subsections of the AM1316-241 dust lane, selected to have different
typical signal-to-noise levels. Indeed, this shows that the perimeters
derived for the original data at transmission 0.9 are overestimated
for smaller signal-to-noise, and that the effect can approach
50\% for the lowest-quality data involved. It is much smaller,
never greater than 20\%, for smoothing scales of two pixels,
and is negligible for 4--8 pixel smoothing. Fig. 7 includes circled points
incorporating these corrections, which fall closer to the linear
extrapolations from larger smoothing scales. There is still some
excess at single-pixel sampling, although the effects of noise are difficult
to model precisely enough to tell whether this is a genuine physical
effect.

There is marginal evidence that the slope of the logarithmic perimeter-scale 
relation (i.e. negative of the box-counting dimension) varies 
systematically with transmission,
from $-0.35$ for $T_I=0.8-0.9$ to $-0.42$ for $T_I=0.5$ (though
the scatter between transmission values is substantial). This
may be due to the overlap of individual structures, or may reflect
intrinsic properties of the distribution. Thus, while the contours
at constant transmission are individually scale-free for sampling
lengths $\ge2$ pixels, the scaling seen for different transmission
values is not consistent. A slope of -0.4 is representative
of the whole data set for this object, leading to a box-counting
dimension $D_b = 1.4$. As noted by Westpfahl et al. (1999),
while the various ways of measuring dimension for arbitrary
contours need not coincide, they are often very close for
astronomical applications. A dimension
$\sim 1.4$ lies within the range 1.2--1.5 found for projected
H I distributions in nearby galaxies by Westpfahl et al.,
so this may be a typical range for cold ISM constituents.

The traditional definition of the fractal dimension comes from
Mandelbrot's equation for a group
of contours at different levels but the same smoothing scale, with 
perimeter $P$ and area $A$:
$$ P^{1/D} = {\rm constant} \times A^{1/2}$$
where the constant may take on different values for various smoothing
scales. For these data, we find that the area enclosed by a contour at a 
given level
is closely constant for various smoothing values, in fact as
constant as numerical scatter will preserve (to within 10\% over
a factor of 8 in smoothing scale for well-sampled contours). 
The near-constancy of area implies that the
mass estimates derived above from absorption at the observed resolution
are not significantly affected by subpixel structure. We are unable to fit
a unique fractal dimension $D$ from this relation (Fig. 8), because the
observed $\log P - \log A$ relation is very nonlinear (being
close to parabolic with a clear maximum). The most we can then say
about the projected dust distribution in this context is that it
has fractal aspects as seen in the scaling behavior of individual
contours, but not as seen in the whole family of contours at 
different levels for a given smoothing length. For large
smoothing scales (4 pixels or more), the scaling behavior exhibits
a slope near 0.35 in the log area - log perimeter plane, implying
a dimension 0.7 (twice the slope).0.7 (twice the slope). This differs
strongly from the box-counting dimension of 1.4, as well as being 
very dependent on smoohting length. The dust distribution we observe
thus shows some aspects of fractal behavior but does not satisfy the
Mandelbot definition for these reasons.

In AM0500-620 the region of overlap is about two times smaller 
than in AM1316-241, so
analysis of its geometry is necessarily sketchier. Applying the box-counting
procedure to a $64 \times 64$-pixel region shows approximate scaling behavior 
all the way down to single-pixel scales (Fig.~9), in contrast to what we see in
AM1316-241. These data are consistent, therefore, with a fractal or
scale-free dust distribution on scales from 50-400 pc, while the 
size of the backlit region does not allow us to measure scales larger
than this. However, we note that higher noise per pixel in this
case might to some extent mimic pixel-scale structure in the contours.
In contrast to AM1316-241, the fitted slope of the
logarithmic perimeter-scale relation flattens on going to larger
extinction, this time from $-0.55$ at $T_I=$0.8--0.9 to 
$-0.2$ at 0.6. This is the level at which the size of the analyzed region
introduces an artificial limit to the perimeter measures.

\subsection{Fine structure and the reddening law}

A major goal of these observations is to quantify the effects of dust on
scales finer than the kpc resolutions afforded by ground-based images.
In particular, we ask whether there are preferred scales for absorbing 
complexes, or whether the spectrum of structure continues down to the WFPC2 resolution
limit.  Our analysis of the structure
in opacity maps, as described above, is one way to address these questions.
In another approach, we consider how the residual (transmitted) intensities 
$T_B$ and $T_I$ are correlated pixel by pixel.  Smoothly distributed
dust and dust clumped in high opacity ``bricks'' are two extremes
for the dust distribution which lead to different predictions for
the wavelength dependence of the extinction.
The extinction of smoothly distributed dust should follow
the grains' intrinsic reddening law (for example, the Galactic extinction curve);
alternatively, if the dust is clumped into opaque clouds, the extinction curve
would be ``grayer'' due to saturation at shorter wavelengths. 
Fig.~10 shows the two-color extinction behavior of the
dust in AM1316-241, for the region with the
highest background intensity and therefore S/N ratio (the same
area used above for the fractal analysis). The error distribution
for individual points will be approximately given by the elliptical scatter
for points near (1,1) in these plots, and this error is not directly dependent
on the local opacity since the background intensity rather than amount
of absorption usually controls the error. This scatter is slightly
correlated between the two passbands 
since fine structure in the foreground galaxy will
consist largely of brighter and bluer stellar associations.
The adjacent panel shows curves predicted for various values 
of $R=A_V/E_{B-V}$, shown for values $R=2.1-4.1$.
Because of the finite passbands and large
extinction range involved, these curves were calculated rigorously 
starting with
the photon-number spectrum of a model old stellar population
(using the models described by Charlot \& Bruzual 1991),
applying reddening using the Cardelli, Clayton, \& Mathis (1989) 
parameterization
and varying the value of $R$, then folding the result through the WFPC2
sensitivity with each relevant filter curve.  To derive a best-fitting
values of the total-to-selective extinction ratio $R$ in the presence of 
comparable relative error in both axes, we consider statistics within
narrow slices of width
0.05 in residual intensity at $B$, and evaluate the mean and median
values of $I$ residual intensity within these slices. The results are
listed, with standard deviations of the means, quartiles of the distributions, 
and modal estimates, in
Table 1, and are 
shown with a family of curves generated for $R=2.1-4.1$ in Fig.~10. 
The listed mode is the IRAF estimator, constructed by automatic binning
and interpolation for such relatively sparse data sets.
These statistics
should be most useful in the middle of the extinction distribution,
where, for example, the tilted cutoff at the high-extinction end
will not bias the results; we considered various ranges in $B$-band
residual intensity for the fit. Using a $\chi^2$ fit to the means (since 
mean values come with well-defined error bars), the range
of residual intensity 0.15-0.90 gives the best fit, reduced $\chi^2 = 2.3$
with one degree of freedom (the value of $R$), namely
$R=3.6 ^{+0.5} _{-0.3}$ where the error range is for 90\% confidence. 
Since variation in $R$ has a very constrained effect on the form of the curve, 
we also considered fits in which the top and bottom bins were included or 
excluded, to test whether their errors, which are largely irreducible 
by changing $R$, dominate the overall statistic.
In all cases, the 90\% confidence bounds on $R$ include the range 3.2--3.6,
with preferred values 3.2--3.6. The $\chi^2$ values for these ranges 
are worse than the fiducial fit above. The value in AM1316-241 is consistent
with the Milky Way value, with some evidence that the extinction curve
is slightly flatter ($R$ greater) than the standard Galactic law. The dust
follows a simple screen model much more closely than we found from
ground-based data (WKC). This must be largely a resolution effect; at
higher spatial resolution, we will see the dust follow its intrinsic
reddening law more closely as clumping becomes less important. There is marginal
evidence for a smaller value of $R$ at smaller extinction, with the formal
fit for the residual-intensity range $0.75-0.9$ of $R=2.9$. However, the number
of fitting bins is small enough to make this at best a tentative statement.

We do see direct evidence for the effective reddening law changing at
different resolution on going to larger scales than the WFPC2 pixels,
which accounts for the greyer extinction found from ground-based
data (White et al. 2000).
Block-averaging the transmission images to lower resolution shows
a transition to a fitted value $R=4.1$ between $0.4\arcsec$ and $0.8\arcsec$ 
pixel sizes. Effective pixel sizes $0.2-0.4\arcsec$  
give color distributions which are consistent with the $R$ value
fit to the original pixels. The transition between $R$ values that appear
representative of the physical grain properties, seen at the full
data resolution, and the artificially grey $R$ values we saw from the
ground, mostly takes place in this range from 200--400 pc resolution.

The color behavior in AM0500-620 is broadly consistent with what we see in
AM1316-241, but less well-determined. The signal-to-noise ratio
of the extinction in each pixel is lower by almost a factor 2, so
the scatter in the two-band transmission plot (Fig.~11) is larger
than for AM1316-241. This is due both to weaker background
light and to a larger contribution from foreground light in the spiral,
which introduces a larger error in correction. A formal $\chi^2$ fit
to statistics in bins of B transmission (Table 2), as done for 
AM1316-241, yields a 90\% confidence interval of $R=1.5 \pm 0.4$, and the
best-fitting $\chi^2 = 2.5$ per degree of freedom, a poor fit. 
At face value, the dust in AM0500-620 has a much larger selective-to-total
extinction ratio than Galactic grain populations. This behavior 
is in the wrong sense to be caused by either scattering or
young stars, unless they are both highly reddened and distributed
in such a way as to show no bright or very red structures in 
color-ratio images. The distribution of $T_I$ in a narrow
range of $T_B$ is quite asymmetric in this object, as seen
from the systematic differences among mean, median, and mode estimates
in Fig.~11. The peak of the distribution, shown by the modal behavior,
most nearly tracks the $R=2-4$ curves, but even in this case the
match is no more than broadly consistent. Repeating the procedure for
only the single dust complex with the best-determined extinction
values (Table 3, Fig.~12) gives results more consistent with the Galactic mean 
for all three measures (mean, median, mode), and a value
$R=2.5 \pm 0.4$, which may indicate that the 
results from the whole overlap region are less trustworthy.

We also examined the two-color reddening behavior in NGC 1275, restricting
the analysis to pixels covering a total area of 55 arcsec$^2$
from 7 rectangular regions including obvious dust features, and
excluding pixels which were masked by a color criterion intended to
reduce contamination from (presumably blue) foreground starlight.
The results are not consistent with a Galactic law (Table 4, Fig.~13).
This could mean that we are unable to remove the foreground light
even with a color criterion, that the dust in the foreground system
of NGC 1275 is quite different from Milky Way dust, or that the
foreground system actually lies well inside the stellar distribution
of NGC 1275 proper. The $R$ parameter becomes ill-determined for
extinction that is so nearly grey; in this instance, values of $R$ in
the range 18--35 are acceptable in a $\chi^2$ sense when fitted to 
the mean values and statistical errors in Table 4.



\section{Summary}

We have used HST imaging of two spiral galaxies partially backlit by
elliptical and S0 systems to provide direct measures of extinction
and dust structure of the foreground spirals. 
This approach provides extinction estimates which are
independent of the assumptions about internal structure that are
necessary to perform many of the statistically-based opacity tests
often applied to this question.

Opacity maps show an interesting contrast between the spiral components of
AM1316-241 and AM0500-620. In both cases, the arm/interarm dichotomy
in extinction is strong. However, the statistics of coverage at
various transmission levels are notably different in the two galaxies.
The arm/interarm dichotomy in extinction is strengthened at the
high resolution afforded by HST; much of what
appeared in an earlier ground-based study as diffuse interarm dust 
is now seen to be comprised of narrow spurs
crossing the main spiral pattern (as in AM0500-620). For the areas
we can analyze, located at radii of $0.4-0.8 ~ R_{25}$, the
dust is heavily concentrated to arms, with half the dust mass 
contained in only 20\% of the projected
area and 95--98\% of the dust mass contained in half the area. 
Kiloparsec-scale complexes of absorbing
material are estimated to contain $3-7 \times 10^4$ $M_{\sun}$ of
dust grains, with correspondingly large implied gas masses,
$1-2 \times 10^7$ $M_{\sun}$, if we assume a typical Galactic gas-to-dust ratio.
Interarm extinction is quite small, although the
statistical distribution by area differs for the two spirals. 
There is a well-defined low-extinction peak in the areal distribution
for AM1316-241, while the distribution of extinction values is much broader
and continuous with the arm regions in AM0500-620.

Reddening curves have been measured by comparing transmitted light 
pixel by pixel between $B$ and $I$ passbands. For AM1316-241, we
find a well-determined slope given by $R = A_V/E_{B-V} = 3.4 \pm 0.2$,
close to the Galactic value of 3.1 derived largely from diffuse 
regions. The data quality for AM0500-620 is poorer, and the overall
fit to a Galactic law (or any plausible reddening law) is poor.
For the single cloud in AM0500-620 with the best data, we do see
reasonable reddening behavior, with a formal fit of $R=2.5 \pm 0.4$.
In NGC 1275, our accuracy is limited by our inability to estimate
the brightness of the foreground galaxy. We have used a color criterion to
eliminate pixels with obvious foreground light and find the remainder
have a reddening curve much flatter (greyer) than the local norm. This might
imply a different dust composition, a different fine-scale distribution,
or that the foreground system is so close to NGC 1275 that it is
at least partially embedded
within its stellar distribution (so that our simple geometric scheme
for measuring opacity breaks down).

We speculate, moreover, that the dust grain size distribution in the NGC 1275
foreground system has been truncated at the low end by sputtering
in the hot intracluster medium of Abell 426. Lack of particles smaller than
$\sim 0.5 \mu$ would flatten the extinction curve effectively.
Following Draine \& Salpeter (1979), the survival time scale
for particles of radius $a$ embedded in a hot medium of density 
$n$ is
$$t_{sp}  =  10^6~{a \over n} ~{\rm years},$$
for $a$ in $\mu$m and $n$ in cm$^{-3}$.
For a representative value of $n = 10^{-3}$, the timescale
for $a = 0.25 \mu$ is $10^9$ years. At the high velocity difference
between this system and NGC 1275 proper (itself essentially at rest in the
cluster) it would have moved at least 3 Mpc during the time it would
take to sputter a Galactic grain population to this extent.
This is comparable to the extent of X-ray emission from the Perseus cluster
(\cite{ettori}).
A strict application of this sputtering prescription would weight
the distribution toward larger grains only for very specific
size distributions, since for spherical grains and a power-law
size distribution, the net effect of sputtering would be to
shrink all grains at a constant rate by area (rather than volume).
In this case, a given set of original grains would move down the
size distribution in step, with the only net affect being progressive
loss of the largest grains. However, recent modeling of sputtering
effects on more realistic grain structures suggest that the net
effect can indeed by a greying of the extinction curve
(\cite{aguirre99a}). Furthermore, preferential ejection of
small grains due to either ram pressure or galactic winds would
weight the distribution toward larger sizes (\cite{aguirre99b}).
If the grain population in the foreground system of NGC 1275
has indeed been altered, the original dust mass and typical $B$
extinction would have been substantially larger than we observe
(with the $I$ extinction being less affected, since larger grains
are more important at this wavelength). If the system has interacted
with the intracluster medium, the small H~I mass (\cite{vge83})
is easier to reconcile with the morphological evidence for a late
Hubble type, since H~I is the first ISM component lost from
cluster spirals (\cite{cayatte}; \cite{bravo}).

We have used the perimeter-smoothing length test to see whether the
dust in these galaxies can be well described by fractal (scale-free)
structures, as has been found for H~I distributions in nearby
spirals. We find scale-free behavior for the perimeter of a given
contour in dust column density as a function
of smoothing scale from 0.05--8 kpc. However, we cannot determine a fractal
dimension from these data, because the scaling between different
contours (at different transmission levels) 
at a given smoothing length does not show the requisite linear
relation for log (area) versus log (perimeter) relation (the perimeter in 
fact peaks within the
measured area range). We can thus describe the dust distribution as
fractal-like but not formally fractal, with box-counting dimension
close to 1.4 but
area-perimeter measures giving a dimensions near 0.7. The box-counting 
dimension is within the range found for H I structures in galaxies by 
Westpfahl et al. (1999), but given the discrepancy between the two
dimensions and the non-monotonic area-perimeter relation, the
dust distribution cannot be accurarely described as fractal. The
area covered at a given extinction level is nearly independent of
smoothing scale over the range we can measure, so
that our dust mass estimates are not biased by structures smaller than
our resolution.

What processes define and control fine structure in the dust? Grains are
generally strongly coupled to other components of the interstellar
medium, both empirically and for robust theoretical reasons, so the
dust distribution will reflect a range of physical processes that need
not impact the grains directly. Strictly dynamical features of disk
dynamics can concentrate the ISM into regions of high density, and
can change the configuration of these regions on dynamical timescales
(a recently discussed example is amplification of small fluctuations
as seen in the nuclear dust patterns of NGC 2207; \cite{elm98}). Energy 
input from stellar winds and supernovae
has a crucial role in altering the small-scale structure of the ISM 
(Rosen \& Bregman 1995), both mechanically and by driving phase changes
(including grain evaporation in the most extreme environments). These
data will not support a detailed analysis of such issues, but the differences
in dust distributions may most plausibly be linked to changes in the 
energy-input rates and perhaps to different
dominant mechanisms shaping the ISM in these two galaxies.

\acknowledgments

This work was supported by NASA HST grant GO-06438.01-95A.
R.~E.~W. was supported in part by a National Research Council 
Senior Research Associateship at NASA GSFC.
We thank Dave Westpfahl for secret tips on doing the fractal-index
measurements within IRAF, and Bruce Elmegreen for asking
some especially probing questions. Wentao Wu provided the specific Bruzual-Charlot
model spectra for rigorous extinction calculations. Alex Rosen pointed
out the role of stellar energy input for the scales of ISM structures.
We thank the referee for forcing us to clarify the role of noise
in the dimensional analysis and actually state the implied
fractal dimensions.

%
%

\clearpage

\figcaption
[keel.f1.eps]
{The F814W image of AM0500-620, displayed with a logarithmic transfer
function to enhance visibility of the dust lanes silhouetted in front
of the northern galaxy. The inset shows the center of the background system
expanded by a factor of four,
illustrating its spiral dust lanes and star clusters. The area shown subtends
$56 \times 62$ arcseconds, and celestial north is $22^\circ$ clockwise from 
the top. All images are shown in positive contrast, to avoid confusion in 
interpreting absorption features.
\label{fig1}}

\figcaption[keel.f2.eps]{F814 image of AM1316-24, shown with a logarithmic intensity
scale. The inset shows the nuclear disk structure in the background elliptical,
at a scale four times larger.
The field shown here is $45 \times 70$ arcseconds from WFPC2
chips 2 and 3, where north is $58^\circ$ clockwise from 
the top. For display purposes, data from the two CCDs have been resampled
to a common astrometric grid using the {\tt wmosaic} task in STSDAS, but 
no such resampling was done for the numerical analysis.\label{fig2}}

\figcaption[keel.f3.eps]
{Transmission maps of the overlap region of AM0500-620 in $B$ and $I$
bands, displayed at the same intensity scale. The region depicted
subtends $10.7 \times 17.5$ arcseconds, with structure near the
nucleus of the background galaxy appearing at the upper left.
The intensity scale bar is linear from zero to 1.5. Two spiral arms
and an interarm dust spur are prominent in absorption. The cloud
examined for reddening behavior in Fig.~8 is just above the center
of each panel.\label{fig3}}

\figcaption[keel.f4.eps]
{Transmission maps of the overlap region of AM1316-241 in $B$ and $I$
bands, displayed at the same intensity scale. Each panel shows
a region $21.0 \times 16.1$ arcseconds. The intensity scale bar is 
linear from zero to 1.5. The apparent edge-on nuclear disk in the
background elliptical galaxy appears at right. There is very
little absorption signature within the prominent and filamentary
spiral feature. \label{fig4}}

\figcaption[keel.f5.ps]
{Transmission maps of the backlit foreground system in NGC 1275, now including
mosaiced parts of all 4 WFPC2 CCDs. The bright star clusters
can be seen closely associated with absorption regions, though the
division by the background model needed to derive transmission
values exaggerates their brightness with increasing distance
from the Seyfert nucleus (which is marked by the white cross in each image).
The intensity scale at the bottom is linear from 0 to 1.5 in transmitted
intensity.
This field is $42.7\arcsec$ across, and north is $48^\circ$ clockwise
from the top. The F450W and F702W filters transmit enough
light from strong emission lines to show some of the low-ionization
filaments in NGC 1275, as seen particularly in the lower left corner 
of each image.\label{fig5}}

\figcaption[keel.f6.ps]
{Distribution of $B$-band extinction by area for regions in AM1316-241 and
AM0500-620. The regions examined are the same ones used for the
fractal analysis, comprising 6617 pixels for AM1316-241 and 3540
pixels for AM0500-620. The smooth curves represent cumulative
distributions, the fraction of the area with less than the indicated
level of transmission. The heavy spiral arm and almost transparent
interarm region make the differential distribution distinctly bimodal 
in AM1316-241, while that in AM0500-620 shows a monotonic (and almost
uniform) decline in affected area with extinction. In each case the
pixel-by-pixel area is typically 0.1, so the number of pixels
below 0.1 (about 0.05) is consistent with measurement error alone. That is, these
data suggest that essentially none of the regions studied have
$A_B > 2.5$ magnitudes. For each panel, the vertical scale refers
to the cumulative distribution, and is relative for the differential
histogram.
 \label{fig6}}

\figcaption[keel.f7.ps]
{Box-counting analysis of the relation between contour length
(region perimeter) and smoothing length,
for a 128-pixel square region in AM1316-241. Contours were
evaluated for values of $I$ transmission 0.5--0.9 as indicated.
Scale-free (fractal) behavior is manifested as straight lines in this
log-log plot, and is a good description of what is seen for smoothing
scales greater than one pixel. The lines shown are best fits to the
middle three points in the
contours at levels 0.5 and 0.9, showing the apparent excess of structure for
no smoothing (single-pixel scales); some of the values
at large smoothing values are likely to be unreliable when small
total areas are involved. This excess amounts to a factor
of two in perimeter length above the extrapolation of the relation
for the 0.9-level contour. The circled points indicate an approximate
correction for finite signal-to-noise in the images, which is
important only for the smallest smoothing scales and largest
transmissions. Much (but not all) of the apparent discrepancy
between unsmoothed and smoothed data is removed by this
correction.\label{fig7}}

\figcaption[keel.f8.ps]
{The defining area-perimeter test for fractal behavior, applied to the
$I$-band extinction in AM1316-241 as sliced at transmission levels
0.5--0.9. A pure fractal distribution will appear as a straight line 
(log-log relation) in this plot, which shows the relation for each
value of smoothing (1--16 pixels) with points indicating values measured
at intervals of 0.1 in transmission (0.9 is at the right). The same
S/N corrections have been made as shown in Fig. 7; the peak and
downturn at small smoothing scales persist even with no such
correction.\label{fig8}}

\figcaption[keel.f9.ps]
{Box-counting analysis of the relation between contour length
(region perimeter) and smoothing length,
for a 64-pixel square region in AM0500-620. Contours were
evaluated for values of $I$ transmission 0.5--0.9 as indicated.
Scale-free (fractal) behavior is manifested as straight lines in this
log-log plot, and is a good description of what is seen for smoothing
scales greater than one pixel. The lines shown are best fits to
each contour level, lying in monotonically increasing order from
0.5--0.9 at the left edge. Any excess above this for single-pixel scales
is marginal, distinguishing this situation from AM1316-241.
The apparent flattening of slopes and crowing of the relation 
for deeper contours toward the bottom of the graph are artifacts
of the small number of available pixels when large smoothing
scales are applied.\label{fig9}}

\figcaption[keel.f10.ps]
{Two-band extinction relation for pixels in the highest-S/N ratio
part of the dust lane in AM1316-241. The upper panel shows the
relation between transmission values at $B$ and $I$ for individual 
pixels, with the predicted curve for a Galactic reddening law and $R=3.1$ 
shown for reference. The lower panel shows mean, median, and mode for
bins of 0.05 in B transmission, along with a family of curves
generated for values of $R=4.1$ (bottom) to $R=2.1$ (top)
in steps of 0.2. Various ranges of transmission all give best-fitting
values in the range $R=3.2-3.6$. In this and similar figures, the lower panel 
is on an expanded scale to emphasize the differences expected for
various values of $R$ and the details of the data distributions.
\label{fig10}}

\figcaption[keel.f11.ps]
{Two-band extinction relation for pixels in the overlapped
region of AM0500-620. The upper panel shows the
relation between transmission values at $B$ and $I$ for individual 
pixels, with the predicted curve for a Galactic reddening law and $R=3.1$ 
shown for reference. The lower panel shows mean, median, and mode for
bins of 0.05 in B transmission, along with a family of curves
generated for values of $R=4.1$ (bottom) to $R=2.1$ (top)
in steps of 0.2. Only the modal values give fits at all consistent
with the Galactic curve.
\label{fig11}}

\figcaption[keel.f12.ps]
{Two-band extinction relation for pixels in the highest-S/N ratio
dust cloud in AM0500-620. The upper panel shows the
relation between transmission values at $B$ and $I$ for individual
pixels, with the predicted curve for a Galactic reddening law and $R=3.1$
shown for reference. The lower panel shows mean, median, and mode for
bins of 0.05 in B transmission, along with a family of curves
generated for values of $R=4.1$ (bottom) to $R=2.1$ (top)
in steps of 0.2. These results are more consistent with
local dust properties (albeit having substantial error bars)
than the wider-area data of Fig.~10, perhaps indicating
that systematics have crept in to the lower-quality measures.\label{fig12}}

\figcaption[keel.f13.ps]
{Two-band extinction relation for pixels that pass our
color criterion to reduce foreground contamination in NGC 1275.
The relation is much greyer than either what was found in the two
AM galaxies or the Galactic mean. The two-color curves are
slightly different here than in the other cases because of the
use of F702W as the $I$-band filter, instead of F814W. Curves are
plotted for fiducial $R$-values to allow direct comparison with
the other systems, at $R = 4.1 \pm 1.0$, as well as the high value $R=25$
in the middle of the acceptable fitting range from $\chi^2$, and the
straight line corresponding to grey extinction ($R = \infty$). 
\label{fig13}}

\clearpage

\begin{table*}
\begin{center}
\begin{tabular}{rrcrrrr}
\multicolumn{7}{c}{$T_I$ Distribution}\\
$T_B$ & Points & Mean        &  Median & 25\% & 75\% & Mode\\
\tableline
0.10 &  91 & $0.408 \pm 0.010$  &   0.397 & 0.340 & 0.463  & 0.309\\
0.15 & 141 & $0.437 \pm 0.007$  &   0.438 & 0.374 & 0.494  & 0.351\\
0.20 & 221 & $0.462 \pm 0.006$  &   0.458 & 0.397 & 0.527  & 0.440\\
0.25 & 228 & $0.512 \pm 0.005$  &   0.519 & 0.453 & 0.564  & 0.549\\
0.30 & 203 & $0.542 \pm 0.006$  &   0.545 & 0.492 & 0.598  & 0.543\\
0.35 & 169 & $0.579 \pm 0.006$  &   0.582 & 0.542 & 0.629  & 0.583\\
0.40 & 163 & $0.614 \pm 0.007$  &   0.611 & 0.562 & 0.665  & 0.613\\
0.45 & 114 & $0.636 \pm 0.008$  &   0.637 & 0.588 & 0.698  & 0.657\\
0.50 &  94 & $0.684 \pm 0.008$  &   0.687 & 0.620 & 0.743  & 0.717\\
0.55 &  86 & $0.736 \pm 0.008$  &   0.748 & 0.678 & 0.796  & 0.760\\
0.60 &  60 & $0.768 \pm 0.012$  &   0.780 & 0.730 & 0.824  & 0.795\\
0.65 &  53 & $0.803 \pm 0.010$  &   0.813 & 0.762 & 0.839  & 0.836\\
0.70 &  52 & $0.814 \pm 0.015$  &   0.812 & 0.781 & 0.861  & 0.813\\
0.75 &  61 & $0.880 \pm 0.008$  &   0.881 & 0.844 & 0.911  & 0.882\\
0.80 &  76 & $0.902 \pm 0.007$  &   0.898 & 0.874 & 0.937  & 0.875\\
0.85 &  72 & $0.927 \pm 0.005$  &   0.918 & 0.900 & 0.952  & 0.909\\
0.90 & 109 & $0.951 \pm 0.005$  &   0.953 & 0.910 & 0.995  & 0.993\\
0.95 & 169 & $0.981 \pm 0.003$  &   0.986 & 0.960 & 1.006  & 0.989\\
\end{tabular}
\end{center}


\tablenum{1}
\caption{
Distribution of $I$-band transmission $T_I$ in slices of $T_B$ for the
dust in AM1316-241 \label{tbl1}}

\end{table*}

\clearpage
\begin{table*}
\begin{center}
\begin{tabular}{rrcrrrr}
\multicolumn{7}{c}{$T_I$ Distribution}\\
$T_B$ & Points & Mean        &  Median & 25\% & 75\% & Mode\\
\tableline
0.35  &   73  & $0.789 \pm 0.019$  &  0.767 & 0.690 & 0.921 & 0.668\\
0.40  &   72  & $0.782 \pm 0.017$  &  0.776 & 0.652 & 0.883 & 0.652\\
0.45  &   92  & $0.816 \pm 0.016$  &  0.804 & 0.703 & 0.913 & 0.743\\
0.50  &  114  & $0.810 \pm 0.013$  &  0.793 & 0.720 & 0.874 & 0.766\\
0.55  &  127  & $0.818 \pm 0.012$  &  0.794 & 0.718 & 0.894 & 0.701\\
0.60  &  148  & $0.862 \pm 0.015$  &  0.827 & 0.745 & 0.942 & 0.816\\
0.65  &  198  & $0.867 \pm 0.015$  &  0.841 & 0.764 & 0.939 & 0.780\\
0.70  &  175  & $0.880 \pm 0.010$  &  0.873 & 0.793 & 0.975 & 0.886\\
0.75  &  222  & $0.881 \pm 0.009$  &  0.865 & 0.787 & 0.959 & 0.850\\
0.80  &  230  & $0.915 \pm 0.008$  &  0.921 & 0.828 & 0.991 & 0.923\\
0.85  &  258  & $0.942 \pm 0.007$  &  0.946 & 0.872 & 1.009 & 1.009\\
0.90  &  295  & $0.950 \pm 0.007$  &  0.960 & 0.902 & 1.017 & 0.996\\
\end{tabular}
\end{center}

\tablenum{2}
\caption{
Distribution of $I$-band transmission $T_I$ in slices of $T_B$ for the
dust in AM0500-620 \label{tbl2}}

\end{table*}

\clearpage
\begin{table*}
\begin{center}
\begin{tabular}{rrcrrrr}
\multicolumn{7}{c}{$T_I$ Distribution}\\
$T_B$ & Points & Mean        &  Median & 25\% & 75\% & Mode\\
\tableline
0.30  &  26     & $0.633 \pm 0.016$ &   0.654  & 0.545 & 0.709 &  0.665\\
0.40  &  34     & $0.666 \pm 0.015$ &   0.636  & 0.607 & 0.701 &  0.631\\
0.50  &  49     & $0.755 \pm 0.013$ &   0.759  & 0.709 & 0.821 &  0.735\\
0.60  &  62     & $0.797 \pm 0.024$ &   0.776  & 0.724 & 0.822 &  0.781\\
0.70  &  75     & $0.851 \pm 0.031$ &   0.829  & 0.778 & 0.878 &  0.847\\
0.80  &  58     & $0.849 \pm 0.011$ &   0.843  & 0.806 & 0.888 &  0.856\\
0.90  &  62     & $0.908 \pm 0.011$ &   0.903  & 0.855 & 0.981 &  0.847\\
\end{tabular}
\end{center}

\tablenum{3}
\caption{
Distribution of $I$-band transmission $T_I$ in slices of $T_B$ for an
isolated dust cloud in AM0500-620 \label{tbl3}}

\end{table*}

\clearpage
\begin{table*}
\begin{center}
\begin{tabular}{rrcrrrr}
\multicolumn{7}{c}{$T_I$ Distribution}\\
$T_B$ & Points & Mean        &  Median & 25\% & 75\% & Mode\\
\tableline
0.40  &  58  & $ 0.469 \pm 0.011 $ & 0.426 & 0.476 & 0.552 &  0.421\\
0.45  & 143  & $ 0.507 \pm 0.006 $ & 0.471 & 0.489 & 0.611 &  0.467\\
0.50  & 208  & $ 0.550 \pm 0.005 $ & 0.516 & 0.549 & 0.648 &  0.505\\
0.55  & 311  & $ 0.601 \pm 0.004 $ & 0.566 & 0.603 & 0.703 &  0.564\\
0.60  & 376  & $ 0.649 \pm 0.004 $ & 0.619 & 0.655 & 0.744 &  0.609\\
0.65  & 516  & $ 0.703 \pm 0.003 $ & 0.671 & 0.709 & 0.797 &  0.669\\
0.70  & 582  & $ 0.751 \pm 0.003 $ & 0.719 & 0.761 & 0.844 &  0.716\\
0.75  & 565  & $ 0.792 \pm 0.002 $ & 0.768 & 0.760 & 0.840 &  0.768\\
0.80  & 583  & $ 0.841 \pm 0.002 $ & 0.818 & 0.798 & 0.873 &  0.782\\
0.85  & 556  & $ 0.885 \pm 0.002 $ & 0.866 & 0.890 & 0.953 &  0.862\\
0.90  & 492  & $ 0.930 \pm 0.002 $ & 0.913 & 0.933 & 0.996 &  0.879\\
\end{tabular}
\end{center}

\tablenum{4}
\caption{
Distribution of $I$-band transmission $T_I$ in slices of $T_B$ for 
absorption regions in NGC 1275 \label{tbl4}}

\end{table*}

\end{document}